\documentclass[journal]{IEEEtran}

\usepackage{cite}
\usepackage{amsmath,amssymb,amsfonts}
\usepackage{algorithmic}
\usepackage{graphicx}
\usepackage{textcomp}
\usepackage{xcolor}
\usepackage{algorithm}
\usepackage{algorithmic}
\usepackage{amsfonts}
\usepackage{framed}
\usepackage{float}
\usepackage{multirow} 

\begin{document}

\title{mSQUID: Model-Based Leanred Modulo Recovery at Low Sampling Rates}

\author{Yhonatan Kvich,~\IEEEmembership{Graduate Student Member,~IEEE,}
        Rotem Arie,
        Hana Hasan,
        Shaik Basheeruddin Shah, ~\IEEEmembership{Graduate Student Member,~IEEE,}
        Yonina C. Eldar,~\IEEEmembership{Fellow,~IEEE}
        
\thanks{Y. Kvich, R. Arie, H. Hasan, and Y.C. Eldar are with the Faculty of Mathematics and Computer Science, Weizmann Institute of Science, Israel.}%
\thanks{S.B. Shah is with the Department of Electrical Engineering, Khalifa University, UAE.}%
\thanks{Corresponding author: Y. Kvich (email: yonatan.kvich@weizmann.ac.il).}%

\thanks{This research was funded by the European Research Council (ERC) under the European Union’s Horizon 2020 research and innovation program (grant No. 101000967-CoDeS and 101119062-SWIMS), as well as by the Israel Science Foundation (grant No. 536/22).
\\S. B. Shah acknowledges support from Khalifa University under the KU-Belgrade joint research collaboration.}

}

\markboth{IEEE Transactions on X,~Vol.~X, No.~X, Month~2025}%
{Author \MakeLowercase{\textit{et al.}}: Title Here}

\maketitle


\begin{abstract}
Modulo sampling enables acquisition of signals with unlimited dynamic range by folding the input into a bounded interval prior to sampling, thus eliminating the risk of signal clipping and preserving information without requiring high-resolution ADCs. While this enables low-cost hardware, the nonlinear distortion introduced by folding presents recovery challenges, particularly under noise and quantization. We propose a model-based deep unfolding network tailored to this setting, combining the interpretability of classical compress sensing (CS) solvers with the flexibility of learning. A key innovation is a soft-quantization module that encodes the modulo prior by guiding the solution toward discrete multiples of the folding range in a differentiable and learnable way. Our method, modulo soft-quantized unfolded iterative decoder (mSQUID), achieves superior reconstruction performance at low sampling rates under additive Gaussian noise. We further demonstrate its utility in a challenging case where signals with vastly different amplitudes and disjoint frequency bands are acquired simultaneously and quantized. In this scenario,  classical sampling often struggles due to weak signal distortion or strong signal clipping, while our approach is able to recover the input signals. Our method also offers significantly reduced runtimes, making it suitable for real-time, resource-limited systems.
\end{abstract}


\section{Introduction}
\label{sec:intro}

Analog-to-digital converters (ADCs) are essential components in digital signal processing systems, transforming analog signals into digital form. However, both the power consumption and cost of ADCs increase with the required sampling rate and dynamic range (DR). To ensure efficient sampling, it is desirable to operate near the lowest permissible sampling rate while maintaining fidelity \cite{eldar2015sampling, mishali2011sub}. According to the Shannon-Nyquist theorem, a bandlimited (BL) signal can be perfectly reconstructed from its uniform samples taken at a rate no less than twice its highest frequency component.

In addition to sampling rate, the ADC’s DR is a critical design consideration. If the amplitude of the input signal exceeds the ADC’s DR, clipping occurs, causing information loss. Traditional approaches to address clipping include oversampling factors (OF) \cite{marks1983restoring, marks1984error}, spectral gap exploitation \cite{abel1991restoring, rietman2008clip}, and dynamic gain control through automatic gain controllers (AGCs) or companders \cite{perez2011automatic, mercy1981review}. These methods, however, come with limitations, such as the need for significant oversampling, precise spectral knowledge, or the introduction of non-linear distortions and bandwidth expansion.

An alternative strategy involves applying a modulo operation to the input signal before sampling, effectively wrapping the amplitude within a fixed DR. This approach, known as modulo sampling or self-reset ADC, has been investigated in various imaging and biomedical contexts \cite{park2007wide, sasagawa2015implantable, yuan2009activity, krishna2019unlimited, zhao2023unlimited, prasanna2020identifiability, zhang2024line}.
The concept of unlimited sampling avoids the need for side information, instead relying solely on modulo samples to reconstruct the original signal \cite{romanov2019above}. Bhandari et al. \cite{bhandari2020unlimited} showed that for BL signals, oversampling beyond the Nyquist rate allows the recovery of higher-order differences (HOD) from modulo samples, enabling signal reconstruction via summation. While effective in noiseless settings, the performance of HOD-based recovery algorithm degrade in the presence of noise, often requiring higher OF for reliable recovery.
To improve robustness and reduce sampling requirements, Azar et al. introduced the beyond bandwidth residual reconstruction ($B^2R^2$) method, which recovers BL signals from modulo samples using time- and frequency-domain separation principles \cite{azar2022residual, azar2025unlimited}. This was further extended by the LASSO-$B^2R^2$ algorithm, which incorporating a sparsity prior on the residuals, resulting in improved efficiency and performance in noisy scenarios \cite{shah2023lasso, shah2024compressed}. This approach maintains robustness while requiring lower computational complexity, making it suitable for practical deployment.

Mulleti et al. \cite{mulleti2024modulo} extended modulo sampling to finite-rate-of-innovation (FRI) signals, demonstrating its applicability beyond the BL setting.
The recovery frameworks developed for both BL and FRI signals have also been validated through hardware prototypes, confirming their feasibility for hardware implementation \cite{mulleti2023hardware, bhandari2021unlimited, florescu2022surprising}.
Existing recovery algorithms assume ideal sampling, ignoring the fact that modulo operations generate high frequencies beyond practical ADC bandwidths, which act as low-pass filters \cite[Sec. 14.3.4]{eldar2015sampling}, \cite{mishali2011sub}. A hardware-aware approach presented in \cite{kvich2025practical} mixes the folded signal before sampling, shifting the spectrum content into the ADC’s operating range and enabling practical recovery.
Further enhancements have emerged from incorporating side information or hardware-accessible fold indicators. Recent results demonstrate that LASSO-$B^2R^2$ performance can be further improved when such binary folding information is available \cite{shah2024compressed}. Bernardo et al. \cite{bernardo2024modulo, bernardo2025modulo} analyzed the method in \cite{shah2024compressed} using 1-bit cross-level indicators, showing improved robustness under quantization for OF above 3. 

While the BL model is widely adopted, many real-world signals exhibit more structure or are better represented in alternative bases. Shift-invariant (SI) spaces provide such models, representing signals as linear combinations of shifted generators \cite{eldar2015sampling, eldar2009compressed, deboor1994structure, christensen2004oblique, aldroubi2001nonuniform, bhandari2011shift}. Recent works \cite{kvich2024Modulo, kvich2024Modulo2} introduced a recovery framework for SI signals under modulo sampling, achieving near-minimal sampling rates with strong reconstruction guarantees. The work in \cite{kvich2025modulo} extended the theory to handle time-varying and potentially unknown folding parameters.

Existing modulo recovery approaches typically formulate the task as an optimization problem. The $B^2R^2$ algorithm expresses recovery as a least-squares (LS) problem on the out-of-band components of the folded signal, while LASSO-$B^2R^2$ leverages the inherent sparsity in the temporal differences of the modulo effect to reformulate the problem as a sparse compressive-sensing (CS) optimization, enabling faster and more accurate recovery.
Algorithm unfolding, originally proposed by Gregor and LeCun \cite{gregor2010learning}, provides a principled framework for transforming iterative optimization algorithms into structured, layer-wise neural networks. 
By combining the interpretability and convergence properties of classical solvers with the flexibility of data-driven learning, unfolding has been successfully applied to a wide range of inverse problems in signal and image processing, including CS, image reconstruction, denoising, and speech enhancement \cite{monga2021algorithm, yang2018admm, li2020efficient, chen2016trainable, revach2022kalmannet, huang2022winnet, shah2024optimization}. The methodology has also gained traction in medical imaging domains such as ultrasound \cite{van2019deep, solomon2019deep} and biomedical imaging \cite{sahel2022deep}, where model-based priors are essential. However, the combination of CS and deep unfolding for addressing the unique challenges of modulo recovery remains largely unexplored.


In this paper, we propose a deep unfolding framework for recovering BL signals from modulo samples. We enhance the existing CS–based approaches for modulo recovery by leveraging unfolding to achieve accurate reconstruction at lower OF and under more challenging noise conditions, while reducing computational complexity.
Building on the LASSO-\(B^2R^2\) formulation, we unroll iterative solvers into a model-based neural network, enabling fast and robust recovery by learning update parameters directly from data. 
To leverage the structure imposed by the nonlinear modulo operation, we incorporate a soft-quantization (SQ) module directly into the unfolded architecture. This module provides a differentiable approximation of the quantization operator introduced in \cite{shlezinger2021deep}, enabling end-to-end gradient-based training. By softly encouraging solutions to align with the discrete folding grid, the SQ module acts as a structural prior that guides the network toward accurate reconstruction.
Our method, modulo soft-quantized unfolded iterative decoder (mSQUID), surpasses existing recovery algorithms under noise and low OFs.
Our method demonstrates strong recovery performance under additive Gaussian noise, outperforming classical ADC sampling in over 90\% for SNRs above 15 dB at an OF of 1.5, and in over 70\% of cases for SNRs above 5 dB at OF = 2. This represents a substantial improvement over existing recovery algorithms, which do not surpass classical sampling performance at OF = 1.5 and require SNRs exceeding 15 dB to do so at OF = 2.

We evaluate our method across a range of signal-to-noise ratio (SNR) and OF values, demonstrating superior accuracy and significantly faster inference compared to \(B^2R^2\) and LASSO-\(B^2R^2\), due to its fixed, shallow network depth. We also consider a challenging scenario involving simultaneous acquisition of weak and strong components with disjoint frequency bands \cite{bernardo2025modulo}. Quantization in this setting leads to significant distortion: a wide DR is needed to avoid clipping the strong signal, but this reduces resolution for the weak one. By applying an analog modulo operation before sampling, we resolve this trade-off using a fixed analog system with a single acquisition channel, while retaining flexibility in the digital recovery. This enables accurate reconstruction and separation of both components, regardless of their spectral content. Our work differs from \cite{bernardo2025modulo} in that we focus on much lower sampling rates, whereas their approach requires oversampling factors above 3 and larger folding parameters.
In the challenging recovery settings, $B^2R^2$ achieves the best reconstruction accuracy, yet our method attains comparable performance with dramatically lower runtimes—often several orders of magnitude faster.

The rest of the paper is organized as follows. Section~\ref{sec:problem_statement} introduces the problem formulation and reviews existing recovery algorithms for modulo sampling. This section also presents a case study focused on the simultaneous acquisition of weak and strong signal components, highlighting the challenges posed by quantization. Section~\ref{sec:unfolding} details the proposed method, mSQUID, including the model-based unfolded network architecture, the data generation process, and the training procedure. Section~\ref{sec:results} provides an evaluation of the method through simulations, comparing its performance to existing techniques under various noise and quantization settings. Finally, Section~\ref{sec:conclusion} concludes the paper and outlines potential directions for future work.

Throughout the paper, we denote the standard second-order (Euclidean), first-order (sum of absolute values), and infinity-order (maximum absolute value) norms of a vector by \( \| \cdot \|_2 \), \( \| \cdot \|_1 \), and \( \| \cdot \|_\infty \), respectively. For matrices, \( \| \cdot \|_2 \) denotes the spectral norm (i.e., the largest singular value).
We denote matrices by bold uppercase letters (e.g., $\mathbf{A}$) and vectors by bold lowercase letters (e.g., $\mathbf{a}$).
The Hermitian (conjugate transpose) of a matrix \( \mathbf{V} \) is written as \( \mathbf{V}^H \), and \( \mathbf{I} \) is the identity matrix. The first-order difference operator is defined as \( \Delta f(n) := f(n) - f(n-1) \), with shorthand \( \hat{f}(n) := \Delta f(n) \). The operator \( \Delta^N \) refers to the \( N \)-th order difference, obtained by applying \( \Delta \) repeatedly \( N \) times. 
The space \( L^2(\mathbb{R}) \) denotes the set of finite energy (square-integrable) functions. For a discrete-time sequence \( a[n] \), its discrete-time Fourier transform (DTFT) is given by \( A(e^{j\omega}) := \sum_{n \in \mathbb{Z}} a[n] e^{-j \omega n} \). For a continuous-time function \( x(t) \in L^2(\mathbb{R}) \), the continuous-time Fourier transform (CTFT) is \( X(\omega) := \int_{\mathbb{R}} x(t) e^{-j \omega t} dt \). We also use the notation \( \mathcal{F}\{x\}(\omega) \) to represent the CTFT. The constant \( e \) denotes Euler's number. We define \( B_{\omega_m} \) as the space of functions whose Fourier-domain support lies within the band \( [-\omega_m, \omega_m] \). Normalized mean squared error (NMSE) is defined as
\begin{equation}
\text{NMSE}(\mathbf{x}, \tilde{\mathbf{x}}) = \frac{\| \mathbf{x} - \tilde{\mathbf{x}} \|_2^2}{\| \mathbf{x} \|_2^2},
\end{equation}
where \(\mathbf{x}\) represents the ground truth and \(\tilde{\mathbf{x}}\) is the prediction.

\section{Problem Statement}
\label{sec:problem_statement}
Consider a BL signal \(f(t) \in L^{2}(\mathbb{R}) \cap B_{\omega_{m}}\) with amplitude that may exceed the ADC's DR of \([- \lambda, \lambda]\). To avoid clipping, we apply an analog non-linear modulo operation to \(f(t)\) before sampling 

\begin{equation}
    f_{\lambda}(t)=(f(t)+\lambda) \bmod 2 \lambda-\lambda.
\end{equation}
The result is a folded signal, \(f_{\lambda}(t)\), with amplitude confined to \([- \lambda, \lambda]\). The folded signal, which is then uniformly sampled at a rate \(\omega_{s} = \frac{2 \pi}{T_{s}}\), where \(T_{s}\) is the sampling interval, resulting in a discrete sequence \(f_{\lambda}\left(n T_{s}\right), n \in \mathbb{Z}\). The sampling rate must be strictly higher than the Nyquist rate of the original signal, i.e., \(\omega_{s} > 2 \omega_{m}\), or equivalently, \(\mathrm{OF} = \frac{\omega_{s}}{2 \omega_{m}} > 1\).

\subsection{Recovery Algorithms}

Our aim is to recover the unfolded samples, $f(nT_S)$, from the modulo samples $f_{\lambda}(nT_S)$ with a robust and computationally efficient algorithm. We denote $f_{\lambda}(nT_S)$ and $f(nT_S)$ as $f_{\lambda}(n)$ and $f(n)$, respectively. To recover the signal, we estimate the residual samples \(z(n)\), defined as the difference between the folded and true samples: \(f_{\lambda}(n) = f(n) + z(n)\), where \(z(n) \in 2 \lambda \mathbb{Z}\). 
Next, we review three existing algorithms for modulo recovery: HOD \cite{bhandari2020unlimited}, $B^2R^2$ \cite{azar2022residual, azar2025unlimited} and LASSO-\(B^2R^2\) \cite{shah2023lasso, shah2024compressed}.


The key idea behind HOD \cite{bhandari2020unlimited} is that the higher-order finite differences of \( f(n) \) decay rapidly due to its BL property,
$
    \| \Delta^N f \|_\infty \leq (T_s \omega_m e)^N \| f \|_\infty.
$
When \( \| \Delta^N f \|_\infty < \lambda \), the modulo operation becomes transparent at the \( N \)-th difference level
$
    \mathcal{M}_\lambda(\Delta^N f(n)) = \Delta^N f(n)
$.

Thus, the algorithm applies the \( N \)-th order difference operator to \( f_\lambda(n) \), unwraps the resulting sequence by rounding each value to the nearest multiple of \( 2\lambda \), and then integrates \( N \) times to recover \( f(n) \). The minimum required order \( N \) is given by
\begin{equation}
    N = \left\lceil \frac{\log \lambda - \log \beta_f}{\log(T_s \omega_m e)} \right\rceil,
\end{equation}
where \( \beta_f \geq \| f \|_\infty \) and is assumed to lie in \( 2\lambda \mathbb{Z} \). HOD guarantees perfect recovery when the sampling interval satisfies \( T_s < \frac{1}{\omega_m e} \), which corresponds to an OF \(> 2\pi e \approx 17 \). While the method tolerates quantization noise under this oversampling condition, it is sensitive to noise, especially at lower sampling rates.

The algorithm $B^2R^2$~\cite{azar2022residual, azar2025unlimited} used two properties of \( f(t) \in L^{2}(\mathbb{R}) \cap B_{\omega_{m}} \) when sampled above the Nyquist rate. 
The first is \textit{time-domain separation}: according to the Riemann–Lebesgue Lemma, \( \lim_{|t| \to \infty} f(t) = 0 \)~\cite{romanov2019above}. This implies that the signal decays in magnitude as \( |t| \) increases. Therefore, there exists a finite index \( N_\lambda \in \mathbb{N} \) such that
\[
|f(nT_s)| < \lambda, \quad \forall |n| > N_\lambda.
\]
In this region, the folded samples equal the original ones: \( f_\lambda(n) = f(n) \). Consequently, the residual sequence \( z(n) = f_\lambda(n) - f(n)=0 \) for all \( |n| > N_\lambda \), meaning it has \textit{finite support} restricted to the interval \( \{-N_\lambda, \ldots, N_\lambda\} \).

The second property is \textit{Frequency-domain separation}: since the signal is sampled above the original signal's Nyquist rate, its spectrum vanishes outside the baseband
\begin{equation}
F(e^{j \omega T_s}) = 0, \quad \omega_m < |\omega| < \frac{\omega_s}{2}.
\end{equation}
From the linearity of the DTFT and using the decomposition \( f_\lambda(n) = f(n) + z(n) \), we obtain
\begin{equation}
F_\lambda(e^{j \omega T_s}) = Z(e^{j \omega T_s}), \quad \omega_m < |\omega| < \frac{\omega_s}{2}.
\end{equation}
Thus, by sampling above the Nyquist rate and examining frequency components outside the original signal's bandwidth, we can extract part of the residual's spectral content.

Combining the time-domain support constraint and frequency-domain expression, we arrive at the following representation:
\begin{equation}
F_\lambda(e^{j \omega T_s}) = \sum_{n=-N_\lambda}^{N_\lambda} z(n) e^{-j n T_s \omega}, \quad \omega_m < |\omega| < \frac{\omega_s}{2}.
\end{equation}
Since $z(n)$ has finite support, its DTFT is a trigonometric polynomial over an interval. A convex optimization problem is formulated:
\begin{equation}
    \begin{split}
        \min_z ||\mathcal{F}_{\mathcal{A}}(f_\lambda - z)||_2^2 \quad
        \text{s.t. } z(n)=0, \, |n|>N_\lambda,
    \end{split}
    \label{eqbbrr}
\end{equation}
where $\mathcal{F}_\mathcal{A}$ denotes the partial DTFT evaluated over $\omega \in \mathcal{A} := (-\omega_s/2, -\omega_m) \cup (\omega_m, \omega_s/2)$. The convex optimization task in (\ref{eqbbrr}) can be solved using the projected gradient descent (PGD) algorithm. Then, the entries are rounded into the nearest values in $2\lambda\mathbb{Z}$.

LASSO-\(B^2R^2\) \cite{shah2023lasso, shah2024compressed} builds on the time-domain and frequency-domain separation properties used in $B^2R^2$, while leveraging the sparsity of the discrete derivative of the residual to enhance the performance of $B^2R^2$.
Since most of the signal’s energy is concentrated in a limited number of samples, we treat $z(n)$ as a finite-length sequence of size $N$, which is chosen large enough to capture this energy concentration.
Applying the first-order difference operator to the difference between the folded and true samples yields
$
    \hat{f}_{\lambda}(n)=\hat{f}(n)+\hat{z}(n)
$.

Note that \( \hat{f}(n) \) is also BL with the same spectral support as \( f(n) \), meaning its DTFT is zero outside the normalized frequency range \( [-\rho \pi,\, \rho \pi] \), where \( \rho = \frac{2\omega_m}{\omega_s} = \frac{1}{\text{OF}} \). Therefore, by applying the DTFT to the derivatives over the frequency interval \( (\rho \pi,\, 2\pi - \rho \pi) \), we obtain

\begin{equation}
\label{eq:F_lam}
    \hat{F}_{\lambda}\left(e^{\frac{j 2 \pi k}{N}}\right) = \sum_{n=0}^{N-1} \hat{z}(n) e^{-j \frac{2 \pi k n}{N}}, \quad \frac{2 \pi k}{N} \in (\rho \pi, 2\pi - \rho \pi).
\end{equation}

In matrix notation
$ \hat{\mathbf{F}}_{\lambda} = \mathbf{V} \hat{\mathbf{z}}, $
where \( \mathbf{V}_{k,n} = e^{-j \frac{2\pi k n}{N}} \), and \( \mathbf{V} \in \mathbb{C}^{M \times N} \) contains rows indexed by frequency indices within the gap \( (\rho\pi, 2\pi - \rho\pi) \).
The authors in \cite{shah2023lasso} showed that the first-order residual \( \hat{\mathbf{z}} \) is sparse, with non-zero entries corresponding to level-crossings of the form \( (2\mathbb{Z} + 1)\lambda \) in \( f(t) \), or \( 2\lambda \mathbb{Z} \) jumps in \( z(n) \).

\begin{figure*}[t!]
    \centering    \includegraphics[width=1\linewidth]{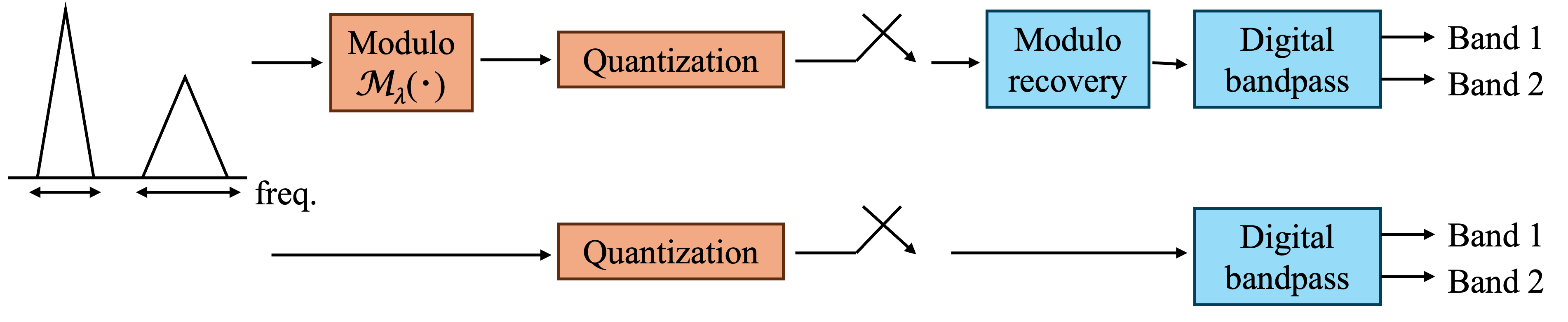}
    \caption{The top illustrates modulo sampling with reduced DR and digital separation, while the bottom shows classical quantization and the same digital separation.}
    \label{fig:high_low_sample}
\end{figure*}
Using this observation, they pose the recovery of \( \hat{\mathbf{z}} \) as a LASSO optimization problem:
\begin{equation}
    \min_{\hat{\mathbf{z}}} \frac{1}{2} \left\| \hat{\mathbf{F}}_\lambda - \mathbf{V} \hat{\mathbf{z}} \right\|_2^2 + \gamma \left\| \hat{\mathbf{z}} \right\|_1,
    \label{eq:lasso}
\end{equation}
where \( \gamma \) is a regularization parameter promoting sparsity.
This is solved via the iterative soft-thresholding algorithm (ISTA), with updates
\begin{equation}
    \hat{\mathbf{z}}^{(i+1)} = \mathcal{S}_{\gamma \tau} \left( \hat{\mathbf{z}}^{(i)} - \tau \mathbf{V}^H ( \mathbf{V} \hat{\mathbf{z}}^{(i)} - \hat{\mathbf{F}}_\lambda ) \right),
    \label{eq:ista}
\end{equation}
where \( \mathcal{S}_\theta(x) = \mathrm{sign}(x) \cdot \max(|x| - \theta, 0) \) is applied element-wise, \( \tau = 1/\|\mathbf{V}\|_2^2 \) is the step size, and \( \gamma = 0.1 \|\mathbf{V}^H \hat{\mathbf{F}}_\lambda \|_\infty \) is used for convergence as suggested in \cite{parikh2014proximal}.
Once \( \hat{\mathbf{z}} \) is recovered, it is rounded to the nearest multiple of \( 2\lambda \), cumulatively summed to obtain \( z(n) \), and the unfolded signal is reconstructed via \( f(n) = f_\lambda(n) - z(n) \).
LASSO-\(B^2R^2\) offers reconstruction accuracy comparable to $B^2R^2$, while reducing computational cost by avoiding iterative sample-wise recovery. 

Each of the algorithms have their own challenges; HOD is computationally efficient but is not robust at lower sampling rates. $B^2R^2$ is computationally expensive, and LASSO-$B^2R^2$ requires high sampling rates for lower $\lambda$ values. This paper introduces a new algorithm, mSQUID, which aims to bridge these gaps.
In the next subsection, we present a specific case study and then continue to describe the deep unfolding approach of our proposed algorithm and outline the training scheme.


\subsection{Simultaneous Acquisition of Weak and Strong Signals}

We examine the acquisition of weak and strong signals in disjoint frequency bands, a setting studied in \cite{bernardo2025modulo}. Classical quantization struggles here: a high-DR ADC avoids clipping the strong component but buries the weak one in noise, while a low-DR ADC preserves the weak signal but distorts the strong.
We model the input as
\begin{equation}
    f(t) = \alpha_1 f_1(t) + \alpha_2 f_2(t),
\end{equation}
where $f_1(t)$ and $f_2(t)$ occupy disjoint bands $\mathcal{F}_1$ and $\mathcal{F}_2 \subseteq [-\omega_m,\omega_m]$ with $\alpha_1 \gg \alpha_2$. Using separate analog bandpass filters is undesirable due to inflexibility and leakage, which can allow energy from $\mathcal{F}_1$ to corrupt the weaker band.

Following \cite{bernardo2025modulo}, we employ a modulo ADC to acquire both signals in a single channel without loss of information, but unlike earlier work we target much lower sampling rates. While the signal is uniquely identifiable above the Nyquist rate of the combined bandwidth \cite{kvich2025modulo}, here we demonstrate through simulations that accurate recovery is achievable even at these lower rates. The modulo operator folds amplitudes into a compact range, mitigating quantization noise dominated by the strong component. The full signal is then recovered with standard modulo algorithms, digitally separated into $f_1(t)$ and $f_2(t)$, and evaluated using NMSE. An overview of the recovery pipelines is shown in Fig.~\ref{fig:high_low_sample}.

\section{Proposed Method: mSQUID}
\label{sec:unfolding}

\subsection{Architecture}

In this section, we present the model-based network architecture \cite{monga2021algorithm}, derived from the ISTA and enhanced with a SQ module to incorporate the prior induced by the modulo operation.
We introduce the following substitution in \eqref{eq:ista}
\begin{equation}
\begin{split}
\mathbf{W}_1 &= \mathbf{I} - \tau \mathbf{V}^H \mathbf{V}, \quad
\mathbf{W}_2 = \tau \mathbf{V}^H.
\end{split}
\label{W_parameter}
\end{equation}
Thus, the ISTA update rule becomes:
\begin{equation}
\hat{\mathbf{z}}^{(i+1)} = \mathcal{S}_{\gamma \tau} \left(\mathbf{W}_1 \hat{\mathbf{z}}^{(i)} + \mathbf{W}_2 \hat{\mathbf{F}}_{\lambda}\right).
\end{equation}
We unfold ISTA into a neural network with $L$ layers, where $L$ is a hyperparameter. In each layer, the trainable parameters are $\mathbf{W}_1^{(i)}$, $\mathbf{W}_2^{(i)}$, and $\gamma^{(i)}$, and the layer-wise update rule is:
\begin{equation}
\hat{\mathbf{z}}^{(i+1)} = \mathcal{S}_{\gamma^{(i)}} \left(\mathbf{W}_1^{(i)} \hat{\mathbf{z}}^{(i)} + \mathbf{W}_2^{(i)} \hat{\mathbf{F}}_{\lambda}\right),
\end{equation}
where $\hat{\mathbf{z}}^{(0)}$ is initialized as a zero vector. 
The weights of $\{ \mathbf{W}_1^{(i)} \}$, $\{ \mathbf{W}_2^{(i)} \}$ and $\{ \gamma^{(i)} \}$ are initialized based on \eqref{W_parameter}, in order to match ISTA before learning begins.
Due to the orthogonality of the Fourier basis, $\mathbf{V}^H \mathbf{V}$ is a scaled identity matrix, i.e., $\mathbf{V}^H \mathbf{V} = N \mathbf{I}$. This allows us to initialize the weight matrices for all iterations as: 
\begin{equation} 
    \mathbf{W}_1^{(i)} = \mathbf{I} - \tau N \mathbf{I}, \quad \mathbf{W}_2^{(i)} = \tau \mathbf{V}^H, 
\end{equation}
where \( \tau = 1/\|\mathbf{V}\|_2^2 \) is chosen to ensure convergence of the ISTA iterations by bounding the Lipschitz constant of the gradient step, which corresponds to the largest eigenvalue of \(\mathbf{V}^H \mathbf{V}\) \cite{parikh2014proximal, shah2023lasso}. The elements of $\mathbf{W}_1^{(i)}$ are real-valued and the elements of $\mathbf{W}_2^{(i)}$ are complex, with each entry of the form:
\begin{equation}
        \left\{\mathbf{W}_2^{(i)}\right\}_{k,n} = \tau e^{\frac{j 2 \pi k n}{N}}.
\end{equation}
To handle the complex values of $\mathbf{W}_2^{(i)}$, we split it into its real and imaginary parts, $\Re\{\mathbf{W}_2^{(i)}\}$ and $\Im\{\mathbf{W}_2^{(i)}\}$, respectively. The update rule becomes
\begin{multline}
\hat{\mathbf{z}}^{(i+1)} = \mathcal{S}_{\gamma^{(i)}} \Big( 
\mathbf{W}_1^{(i)} \hat{\mathbf{z}}^{(i)} 
+ \Re\{ \mathbf{W}_2^{(i)}\} \Re\{\hat{\mathbf{F}}_{\lambda}\} \\
- \Im\{\mathbf{W}_2^{(i)}\} \Im\{\hat{\mathbf{F}}_{\lambda}\} 
\Big)
\end{multline}
where we note that $\hat{\mathbf{z}}$ is real-valued, so only the real part of the update is retained.
\begin{figure*}[htb] 
    \centering
\includegraphics[width=1\linewidth]{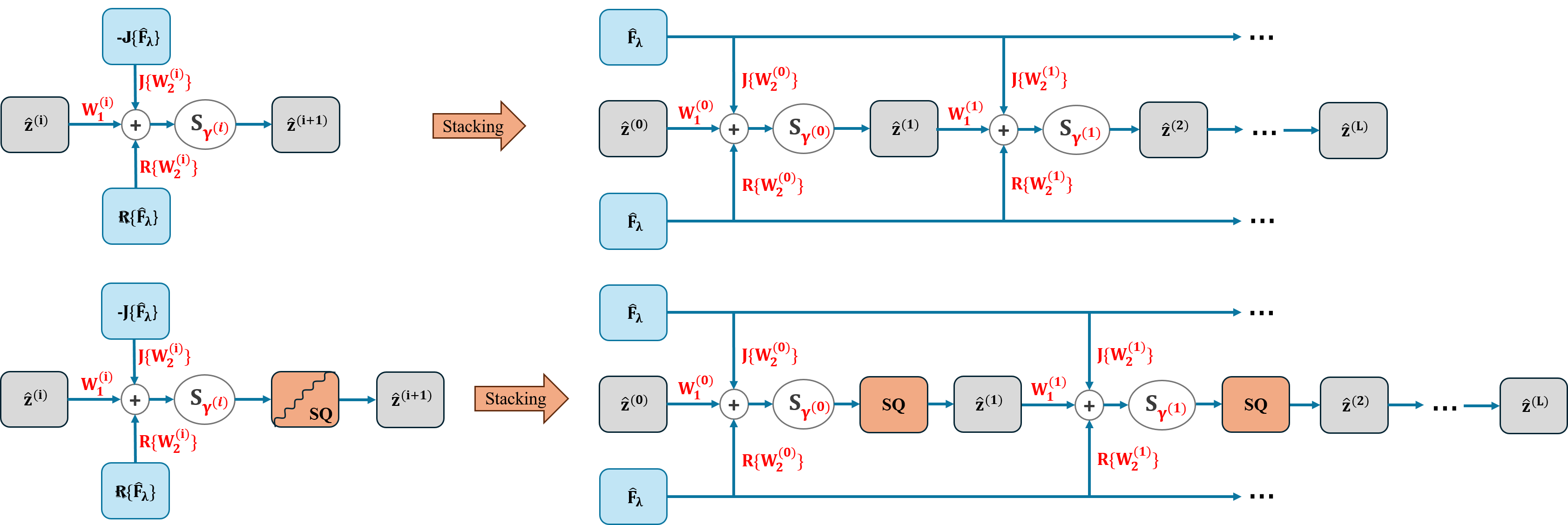}
    \caption{A single network layer and the unfolded deep model-based network, shown (a) without and (b) with a SQ step.} 
    \label{fig:lista}
\end{figure*}

We now introduce an additional step designed to incorporate the structural prior imposed by the nonlinear modulo behavior. Specifically, in the modulo folding of the input signal values are effectively mapped to the nearest multiple of $2\lambda$. This motivates the inclusion of a quantization-like operation that enforces this structural prior. To model this behavior in a differentiable way, we adopt a SQ operator based on a sum of hyperbolic tangent functions \cite{shlezinger2021deep}
\begin{equation}
Q_{\text{soft}}(x) = \sum_{k=1}^{D}a_k \cdot \tanh(c_k x - b_k),
\end{equation}
where $\{a_k\}$, $\{b_k\}$, and $\{c_k\}$ are learnable real-valued parameters and $D$ is the number of quantization levels. This formulation resembles a piecewise step function that is differentiable, while allowing gradient-based optimization.

We adapt this formulation into a custom SQ tailored to our setting. Recall that $\mathbf{z}$ captures the modulo correction applied to keep the signal within $[-\lambda, \lambda]$, and its discrete derivative $\hat{\mathbf{z}}$ indicates when these corrections occur. Each such correction corresponds to shifting the signal by an integer multiple of $2\lambda$, so the entries of $\hat{\mathbf{z}}$ lie in $2\lambda\mathbb{Z}$. To exploit this structure, we introduce a quantization-like operator with levels fixed to this grid. Instead of performing exact rounding to the nearest element of $2\lambda\mathbb{Z}$, which is non-differentiable, our SQ module provides a differentiable and learnable approximation, enabling gradient-based training while incorporating the modulo prior.
The SQ in our case is given by
\begin{equation}
Q_{\text{soft}}(x; \beta) = -\lambda + \sum_{i=1}^{D} \lambda \cdot \tanh\left( \beta (x - s_i) \right),
\end{equation}
where $\beta$ controls the steepness of transitions, a larger value leads to a decrease in steepness approximations of hard quantization, and the shift points $s_i$ are
\begin{equation}
s_i = (i - (\ell + 1)) \cdot 2\lambda, \quad \text{s.t  } D=2\ell+1.
\end{equation}
The update rule within a single fold of the network that includes the SQ step becomes
\begin{multline}
    \hat{\mathbf{z}}^{(i+1)} = Q_{\text{soft}}\Big(\mathcal{S}_{\gamma^{(i)}} \big(\mathbf{W}_1^{(i)}\hat{\mathbf{z}}^{(i)} + \Re\{\mathbf{W}_2^{(i)}\} \Re\{\hat{\mathbf{F}}_{\lambda}\} -\\  \Im\{\mathbf{W}_2^{(i)}\} \Im\{\hat{\mathbf{F}}_{\lambda}\} \big) ; \beta^{(i)}\Big).
\end{multline}

This construction allows every input value to be softly ``pulled" toward its nearest $2\lambda \mathbb{Z}$ level. The number of levels $D$ is selected to cover the full DR of the input signal. $\beta$ is learned for each layer. The steepness parameter $\beta$ also serves to control how ``harsh" the SQ is at each layer, giving the model flexibility to adapt the quantizer during recovery. This structure proved beneficial in prior work \cite{shlezinger2021deep}, where SQ was used to train models that are ultimately deployed with hard quantization. Instead of training directly on pre-quantized samples, the SQ was learned jointly with the network and later replaced with its hard counterpart, leading to improved recovery and better alignment with real hardware constraints. 
Fig.~\ref{fig:lista} depicts the unfolded network architecture with and without the inclusion of SQ.

\subsection{Data}
We generate datasets for training and evaluation, grouped into two scenarios: one with additive white Gaussian noise (AWGN) and another for combined weak and strong signal case under quantization.
In the first class of datasets, we generate 6000 distinct BL signals for each combination of OF $\in\{1.5, 2, 2.5, 3\}$, SNR $\in \{0, 5, 10, 15, 20, 25\}$ dB and modulo threshold is set as $\lambda = 0.25$. Each set is randomly split into 5000 signals for training and 1000 signals for testing. Each BL signal is of length $N = 1024$ and is normalized to a maximum amplitude of one. A modulo operation with respect to $\lambda$ is applied to each signal. AWGN is added at the specified SNR levels, with each signal receiving a noise level varied uniformly by $\pm5$\,dB to reflect practical conditions where the exact SNR is not known in advance.
The resulting noisy modulo-folded signal is denoted \( f_{\lambda}^{*}(n) \),  and its first-order difference is $\hat{f}_{\lambda}^{*}(n)$. The DTFT of the frequency components outside of $[-\rho\pi, \rho\pi]$ of $\hat{f}_{\lambda}^{*}(n)$ is denoted $\hat{\mathbf{F}}_{\lambda}^{*}$, as defined in equation (\ref{eq:F_lam}).

The second dataset is constructed for a case study of simultaneous acquisition of weak and strong signals sampled using low number of bits. It includes another 6000 BL signals with a 5000-1000 training-testing split. Here, the OF and $\lambda$ parameters are kept within the following ranges: OF $\in \{1.5, 2, 2.5, 3\}$ and $\lambda\in\{0.2, 0.25\}$. Each signal is synthesized by combining low-frequency components with strong high-frequency components. The low frequency band is $[-20, 20]$ Hz range, and the high frequency band is modulated by $70$ Hz to the $[50, 90]$ Hz range. An amplification ratio of 1:4 is applied by setting $\alpha_1=1.0$ and $\alpha_2=0.25$. The composite signals are normalized and undergo modulo folding with respect to $\lambda$. Instead of AGWN, signal quantization is introduced by simulating a 4-bit ADC. 
The quantization levels are uniformly spaced within the DR of the ADC.
We use the same notation as in the AWGN case to denote the quantized signal, its first-order difference, and the DTFT beyond the Nyquist interval of the combined signal. In this context, frequencies beyond $90$ Hz, which is the maximum frequency in the combined input signal.

\subsection{Training}

We train a separate neural network for each configuration defined in the datasets described above. 
Each network is trained on a dataset consisting of 5000 training samples. The input to the network is the DTFT of the first-order difference of the noisy modulo signal, denoted as $\hat{\mathbf{F}}_{\lambda}^{*}$, and the target output is the first-order difference of the true residual signal, denoted as $\hat{\mathbf{z}}$.
Specifically, \( \hat{\mathbf{F}}_{\lambda}^{*} \) contains the frequency components of the DTFT of the noisy modulo signal within the interval \( (\rho\pi, 2\pi - \rho\pi) \), which lies outside the original signal's spectral support, as described in \eqref{eq:F_lam}.

During training, the network learns the parameters $\{\gamma^{(i)}, \mathbf{W}_1^{(i)}, \mathbf{W}_2^{(i)}, \beta^{(i)}\}_{i=1}^L$ using backpropagation and the ADAM optimizer \cite{kingma2014adam}. The loss function used is the NMSE.
To determine the best hyperparameters, we employ a Bayesian optimization framework using Tree-structured Parzen Estimators (TPE) \cite{akiba2019optuna}. The hyperparameter search space includes the number of layers $L \in \{4, 5, \dots, 12\}$, threshold initialization $\gamma \in [ 10^{-5}, 10^{-1}]$, and learning rate $\in [10^{-5}, 10^{-2}]$. SQ module introduces its own learnable parameter: the sharpness parameter $\beta$, which controls the steepness of the approximation and is independently set for each layer. To evaluate the impact of the SQ module, we conducted three training runs: one without SQ (equivalent to a standard LISTA network for this task), one with SQ included, and one where a binary hyperparameter flag determined whether the SQ module was used, allowing the architecture search to optimize over this choice. Each TPE trial consists of a full training run for 200 epochs with different hyperparameter configuration drawn from the search space, followed by evaluation based on validation NMSE. We perform 20 such trials and select the best-performing hyperparameter set. To account for randomness in training, we repeat the optimization multiple times and retain the best overall trial per dataset. The final model is then trained on the training dataset using the selected hyperparameters for 200 epochs before testing.

For the full modulo recovery process, the output $\hat{\mathbf{z}}^{(L)}$ is rounded to the nearest multiple of $2 \lambda \mathbb{Z}$ to enforce integer-valued solutions:
\begin{equation}
    \hat{\mathbf{z}}^{(L)} \leftarrow 2\lambda\left\lceil \frac{\lfloor \hat{\mathbf{z}}^{(L)} / \lambda \rfloor}{2} \right\rceil.
\end{equation}
Finally, a cumulative summation is applied to reconstruct the estimated residual signal $\tilde{z}(n)$, and the estimated signal $\tilde{f}(n)$ is computed as
\begin{equation}
    \tilde{f}(n) = f_{\lambda}^{*}(n) - \tilde{z}(n).
\end{equation}
Although the raw signal $f_{\lambda}^{*}(n)$ is not directly used as input to the network, we used the derived $\hat{\mathbf{F}}_{\lambda}^{*}$ as the network's input. The full time-domain signal $f_{\lambda}^{*}(n)$, however, is crucial for reconstructing the final output. The summary of the proposed recovery is presented in Algorithm~\ref{alg:proposed_alg}.

\begin{algorithm}[H]
\centering
\begin{algorithmic}[1]
\STATE \textbf{Input:} Folded samples $f^{*}_{\lambda}(n)$ (corrupted by either additive Gaussian noise or quantization), folding parameter $\lambda$, OF, number of samples $N$, and either the SNR (Gaussian case) or the number of quantization bits (two-component case).
\STATE Compute $\hat{F}_{\lambda}\!\left(e^{\tfrac{j 2 \pi k}{N}}\right)$ for all $k$ with $\tfrac{2 \pi k}{N} \in (\rho\pi,\,2\pi-\rho\pi)$.
\STATE Select the mSQUID network trained for the corresponding noise model (Gaussian with given SNR, or number of quantization bits) and estimate $\hat{\mathbf{z}}$.
\STATE $\hat{\mathbf{z}}\leftarrow 2\lambda\Bigl\lceil \dfrac{\lfloor \hat{\mathbf{z}}/\lambda \rfloor}{2} \Bigr\rceil \quad\triangleright$ round to the nearest element of $2\lambda\mathbb{Z}$.
\STATE $\mathbf{z}\leftarrow \operatorname{cumsum}\!\bigl(\hat{\mathbf{z}}\bigr)\quad\triangleright$ cumulative summation.
\STATE $f(n)\leftarrow f^{*}_{\lambda}(n)-z(n)\quad\triangleright$ reconstruct unfolded signal.
\IF{two-component case with disjoint frequency bands and quantized input}
    \STATE Apply digital bandpass filtering to $f(n)$ to extract the individual components in each frequency band.
\ENDIF
\STATE \textbf{Output:} Recovered signal $f(n)$ (and separated components if applicable).
\end{algorithmic}
\caption{mSQUID}
\label{alg:proposed_alg}
\end{algorithm}

\section{Experimental Results}
\label{sec:results}

In this section, we evaluate the performance of our unfolded algorithm in comparison to HOD, $B^2R^2$, and LASSO-$B^2R^2$ under two scenarios: Gaussian noise and the proposed weak and strong signal case study. All methods are tested on a dataset of 1000 signals. 
As part of the architecture optimization process (using TPE), we explored network configurations both including and excluding the SQ module. In most scenarios, the highest-performing models incorporated the SQ module, highlighting its effectiveness in leveraging the modulo prior during training.
For the Gaussian noise setting, we compare against a classical ADC with AGWN noise matching the specified SNR. For the weak and strong case study, the baseline is a standard uniform quantizer with the same number of bits used in the modulo sampling setup. In both cases, we report the percentage of signals for which each method achieves a lower NMSE than the corresponding classical ADC baseline. This will be called outperformance rate.
Different settings were trained using various GPU models, including NVIDIA A40, NVIDIA L40S, and Quadro RTX 8000. Across both the Gaussian noise and case study scenarios, the complete training process---including architecture exploration via TPE---took no more than 7 minutes per setting. This highlights the efficiency of our approach and its practicality for rapid experimentation and deployment.

\begin{figure*}[htb]
    \centering    \includegraphics[width=1\linewidth]{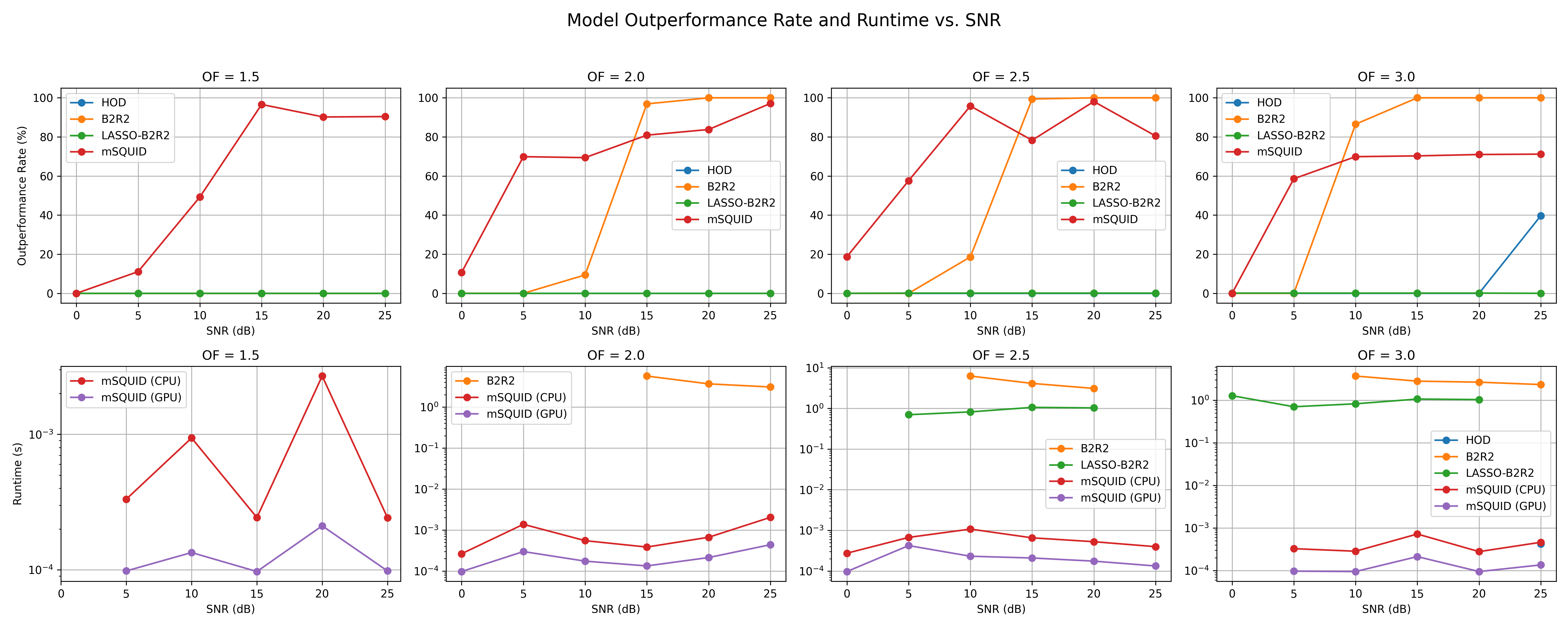}
    \caption{Outperformance Rate (\%) relative to a classical ADC versus SNR, and average runtime versus SNR, for $\text{OF} \in {1.5, 2, 2.5, 3}$ with $\lambda = 0.25$. Runtime results are shown only when the outperformance rate is nonzero. In many cases, the HOD curve (blue) is not visible because it coincides with other curves at zero.}
    \label{fig:results}
\end{figure*}
\begin{figure*}[htb]
    \centering    \includegraphics[width=1\linewidth]{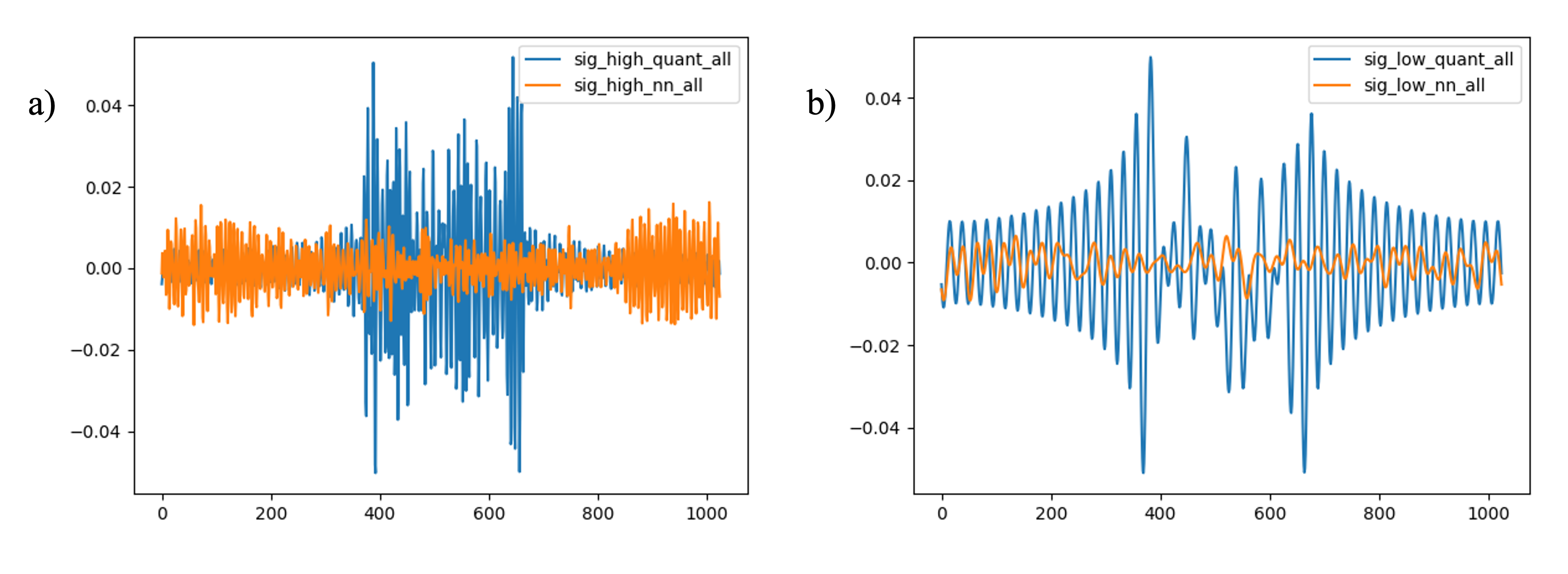}
    \caption{Comparison between classical sampling and mSQUID recovery for a signal with disjoint high and low frequency components (OF = 2.5). (a,b) Reconstruction error on the high and low frequency bands, respectively, for an example test signal. mSQUID achieves lower error, especially where the original signal has high amplitude.}
    \label{fig:high_low_exmple}
\end{figure*}

The outperformance rates under AGWN are shown in the top row of Fig.~\ref{fig:results}.
We first consider a low OF of 1.5, as shown in the top-left subfigure of Fig.~\ref{fig:results}. In this setting, none of the classical recovery methods are able to outperform the classical ADC baseline, whereas our unfolded network surpasses the ADC in more than 80\% of the test cases once the SNR exceeds 10\,dB.
At SNR levels of 15 and 25 dB, the best performances are achieved by the run that includes the SQ module, highlighting its effectiveness in improving reconstruction quality.
As OF increases, the recovery rate improves as expected. At OF\,=\,2, our method performs particularly well at 5 and 10\,dB where SQ module was used and reaches near-perfect recovery at 25\,dB---matching the performance of $B^2R^2$. For other SNR, our network lags behind $B^2R^2$, highlighting that the effectiveness of data-driven training can vary across different sampling and noise conditions. At OF\,=\,2.5, our method leads for SNR\,$\leq$\,10\,dB, while $B^2R^2$ surpasses it at higher SNRs.
At OF=3, our network outperforms the classical ADC in approximately 60\% of the cases at 5 dB, demonstrating an advantage in low-SNR regimes. At higher SNR levels, $B^2R^2$ achieves better reconstruction accuracy, but at the cost of substantially longer runtimes. 

To assess computational efficiency, we compared runtimes of all methods on the same test set. Our architecture uses a fixed and small number of iterations, enabling fast inference. To demonstrate that this speed advantage stems from the algorithm's structure rather than hardware acceleration, we benchmarked both GPU and CPU-only versions of our method. Runtime comparisons are shown in the bottom row of Fig.~\ref{fig:results}, limited to cases where our method achieved a non-zero outperformance rate. While $B^2R^2$ achieves strong reconstruction in many regimes, it requires many iterations and takes several seconds per signal. In contrast, our method is approximately 1000 times faster even on CPU, and significantly faster with GPU acceleration. This makes our network well-suited for scenarios where runtime is critical, balancing performance and computational efficiency.

\begin{table*}[ht]
\centering
\caption{Outperformance rate (\%), NMSE improvement (dB), and runtime metrics for the weak--strong signal case study using 4-bit quantization. Results are reported for $\lambda = 0.25$ and $\lambda = 0.2$. NMSE improvement is reported for the full signal bandwidth, as well as separately for the high and low frequency bands.
}
\begin{tabular}{|c|c|c|c|c|c|c|c|c|}
\hline
$\lambda$ & OF & Method & Outperformance & All Band & Low Band & High Band & CPU & GPU \\
 &  &  & Rate (\%)& Improvement (dB $\uparrow$) & Improvement (dB $\uparrow$) & Improvement (dB $\uparrow$) & Runtime (s) & Runtime (s) \\
\hline
\multirow{16}{*}{0.25}
 & 1.5 & HOD & 0 & - & - & - & - & - \\
 &  & $B^2R^2$ & 0 & - & - & - & - & - \\
 &  & LASSO-$B^2R^2$ & 0 & - & - & - & - & - \\
 &  & mSQUID & 0& - & - & - & - & - \\
\cline{2-9}
 & 2.0 & HOD & 0 & - & - & - & - & - \\
 &  & $B^2R^2$ & 36.6 & 6.74 & 7 & 6.12 & 12.32 & - \\
 &  & LASSO-$B^2R^2$ & 6.9 & 6.6 & -22 & -23 & 1.02 & - \\
 &  & mSQUID & 0& - & - & - & - & - \\
\cline{2-9}
 & 2.5 & HOD & 51.2 & 9.26 & 8.01 & 6.06 & 0.00098 & - \\
 &  & $B^2R^2$ & 99.6 & 7.31 & 8.07 & 6.23 & 9.12 & - \\
 &  & LASSO-$B^2R^2$ & 74.1 & 7.29 & -1.92 & -4.2 & 0.35 & - \\
 &  & mSQUID & 85.6 & 8.66 & 9.7 & 7.55 & 0.013403 & 0.015591 \\
\cline{2-9}
 & 3.0 & HOD & 100 & 10.26 & 9.01 & 6.53 & 0.000474 & - \\
 &  & $B^2R^2$ & 100 & 7.82 & 9.01 & 6.53 & 2.17 & - \\
 &  & LASSO-$B^2R^2$ & 99.9 & 7.82 & 9.01 & 6.53 & 0.084137 & - \\
 &  & mSQUID & 99 & 8.86 & 10.21 & 7.58 & 0.009764
 & 0.010571
 \\
\hline
\multirow{16}{*}{0.2}
 & 1.5 & HOD & 0 & - & - & - & - & - \\
 &  & $B^2R^2$ & 0 & - & - & - & - & - \\
 &  & LASSO-$B^2R^2$ & 0 & - & - & - & - & - \\
 &  & mSQUID & 0& - & - & - & - & - \\
\cline{2-9}
 & 2.0 & HOD & 0 & - & - & - & - & - \\
 &  & $B^2R^2$ & 9.5 & 8.76 & 9.78 & 7.91 & 14.52 & - \\
 &  & LASSO-$B^2R^2$ & 0.1 & 11.46 & -23.46 & -27.28 & 1.27 & - \\
 &  & mSQUID & 0 & - & - & - & - & - \\
\cline{2-9}
 & 2.5 & HOD & 0 & - & - & - & - & - \\
 &  & $B^2R^2$ & 98.4 & 9.47 & 10.87 & 8.3 & 9.79 & - \\
 &  & LASSO-$B^2R^2$ & 21.3 & 9.42 & -4.67 & -9.12 & 0.55 & - \\
 &  & mSQUID & 52.7 & 9.79 & 10.87 & 8.37 & 0.013177 & 0.016225
 \\
\cline{2-9}
 & 3.0 & HOD & 33.6 & 12.65 & 11.7 & 8.47 & 0.00057 & - \\
 &  & $B^2R^2$ & 100 & 9.93 & 11.9 & 8.65 & 5.02 & - \\
 &  & LASSO-$B^2R^2$ & 86.6 & 9.94 & 11.87 & 8.58 & 0.311 & - \\
 &  & mSQUID & 95.4 & 10.67 & 12.78 & 9.41 & 0.011071
 & 0.013591 \\
\hline
\end{tabular}
\label{tab:merged_lambda_table}
\end{table*}

Table~\ref{tab:merged_lambda_table} presents the results for the weak and strong case study. As in the Gaussian case, we report the outperformance rate, defined as the percentage of signals for which each method achieves lower NMSE than the corresponding classical ADC baseline. When outperformance is observed, we also report the average NMSE improvement (in dB), both over the entire signal bandwidth and within the specific frequency bands of the weak and strong components (after digital bandpass filtering). Runtime comparisons are included for all methods, with CPU and GPU results shown for our approach. As before, we omit NMSE improvement and runtime results when the outperformance rate is zero.

For both $\lambda = 0.25$ and $\lambda = 0.2$, meaningful outperformance is observed only at higher OF = 2.5 and 3. The best-performing networks were obtained from runs where the SQ module was always enabled, outperforming those trained without it and highlighting the benefit of incorporating the SQ module. At OF = 2.5, $B^2R^2$ consistently achieves near-perfect recovery, while our method performs slightly below it, outperforming HOD and LASSO-$B^2R^2$ in both recovery rate and NMSE. In successful cases, our method yields an NMSE improvement of around 1\,dB in both the full signal and individual bands. At OF = 3.0, HOD achieves perfect recovery for $\lambda = 0.25$ with very low runtime due to its lightweight structure. For $\lambda = 0.2$, our method achieves a $95.4\%$ outperformance rate, with $B^2R^2$ still recovering all signals, while other methods remain far behind. In this case, our approach and $B^2R^2$ represent a trade-off between runtime and reconstruction accuracy.

Across all settings, our method offers significantly faster runtimes, demonstrating its practicality under low-latency constraints. In the Gaussian noise experiments, GPU inference was consistently faster, even with a batch size of one, since GPUs leverage intra-operation parallelism in matrix computations. By contrast, in the weak–strong case study, all methods required longer runtimes, and the gap between CPU and GPU execution narrowed. In some cases, the CPU was slightly faster due to GPU memory-transfer overhead outweighing the parallelism gains.

Fig.~\ref{fig:high_low_exmple} illustrates presents a recovery example of the case study and compare the performance of modulo sampling using our mSQUID method against a classical sampling approach. The example shown corresponds to OF = 2.5, and we applied the mSQUID network trained for this setting. The signal used in the figure is taken from the test set.
Subfigures (a) and (b) present the reconstruction errors for the same test signal, pandpass on the high-frequency and low-frequency bands, respectively. The results compare the classical sampler (blue) with our mSQUID method (orange). Our method consistently achieves lower reconstruction errors across the entire signal duration, especially in the middle region where the original input signal has high amplitude. In this region, the classical sampler suffers from significant quantization error due to limited resolution over a large DR, whereas modulo sampling maintains high fidelity by compressing the DR prior to quantization.

\section{Conclusion}
\label{sec:conclusion}



We present a model-based deep unfolding network for recovering signals from modulo samples, building upon optimization-based formulations such as LASSO for modulo recovery. This framework unrolls these iterative algorithms into a compact, learnable architecture that preserves the interpretability of model-based methods while enabling fast, data-driven inference.
A key innovation is the SQ module, which embeds the prior imposed by the modulo operation: during training, each estimate is softly ``pulled” toward integer multiples of $2\lambda$ in a differentiable, learnable fashion. This addition strengthens the model-based structure and improves reconstruction accuracy without sacrificing gradient flow.
Experiments confirm the benefits of the proposed approach. Under Gaussian noise, our network surpasses classical ADCs and state-of-the-art algorithms at low OF and in noisy conditions.

In a more challenging setting, where two signals with disjoint frequency bands and vastly different amplitudes are combined and measured under quantization constraints, our method achieves accuracy comparable to the best-performing $B^2R^2$ algorithm in many cases, while offering much faster runtimes—often by more than three orders of magnitude. For the simpler case of higher oversampling (OF = 3) and moderate folding range ($\lambda = 0.25$), the HOD method attains the best performance and runtime; however, this represents the least challenging scenario. This highlights a favorable trade-off between reconstruction quality and computational latency, making the approach well-suited for real-time and resource-constrained systems.
These results establish deep unfolding with SQ as a practical and scalable solution for nonlinear sampling. Future work will explore extensions to additional signal classes, multidimensional data, and task-specific applications in imaging, communications, and sensing.

\bibliographystyle{IEEEtran}
\bibliography{refs}

\end{document}